# XCloud-pFISTA: A Medical Intelligence Cloud for Accelerated MRI

*Yirong Zhou, Chen Qian, Yi Guo, Zi Wang, Jian Wang, Biao Qu, Di Guo, Yongfu You, Xiaobo Qu*\**

*Abstract* — Machine learning and artificial intelligence have shown remarkable performance in accelerated magnetic resonance imaging (MRI). Cloud computing technologies have great advantages in building an easily accessible platform to deploy advanced algorithms. In this work, we develop an open-access, easy-to-use and high-performance medical intelligence cloud computing platform (XCloud-pFISTA) to reconstruct MRI images from undersampled k-space data. Two state-of-the-art approaches of the Projected Fast Iterative Soft-Thresholding Algorithm (pFISTA) family have been successfully implemented on the cloud. This work can be considered as a good example of cloud-based medical image reconstruction and may benefit the future development of integrated reconstruction and online diagnosis system.

## I. INTRODUCTION

MAGNETIC resonance imaging (MRI) is a significant clinical diagnosis tool but suffers from relatively long data acquisition time. To achieve fast imaging, sparse sampling and parallel imaging are hot topics for decades [1-7]. Optimization methods and deep learning have raised great concerns for accelerating imaging speed and improving image quality. The former usually takes the model reconstruction as an inverse problem, which is solved through continuous iteration until convergence [8-11]. The latter trains deep neural network with a big database to reconstruct images [12-14]. Although both of them have excellent performance, the environment configuration is complex and cumbersome. To the best of our knowledge, there are few integrated and easy-to-use platforms for MRI researchers to easily and effectively do the medical image processing and analysis.

Cloud computing is a data-centered distributed computing mode. It is accessible, high-integrated, safe and has virtualized hardware [15, 16]. Compared with traditional computing mode, it has superior technologies in data storage and computing power. Also, users could conveniently utilize the resource on the cloud once login the platform. Moreover, the cloud computing platform provides a unified environment for various methods, which frees users from complex and tedious environment configuration. We are committed to developing the intelligent cloud computing platform for MR signal processing and analysis. In our previous work, several advanced approaches for accelerated Nuclear Magnetic Resonance (NMR) spectroscopy have been deployed on the open access cloud platform successfully[17, 18]. So we believed that, it could be very valuable for medical imaging to deploy various state-of-the-art MRI reconstruction approaches on the cloud.

In this work, we developed XCloud-pFISTA, an easy-to-use, high-performance, and open access medical intelligent cloud for accelerated MRI. So far, two representative MRI reconstruction approaches have been implemented on the cloud: 1) Projected fast iterative soft-thresholding (pFISTA) [19], an model-based optimization method for single-coil MRI reconstruction; 2) pFISTA-Net [20], a model-driven deep learning for parallel MRI reconstruction. Users can enjoy these advanced approaches to conveniently and efficiently recover high-quality images from partial sampled k-space.

## II. METHODS

### A. MRI Reconstruction Approaches

MRI reconstruction from undersampled data by exploiting sparsity of MRI images and low-rankness of k-space raises great concerns[3, 8]. Qu *et al.* proposed pFISTA, an iterative approach for single-coil MRI reconstruction, which employs sparsity of images under the redundant representation of tight frame[19]. The guaranteed convergence analysis of the parallel imaging version was given in the subsequent work[10]. Compared with other state-of-the-art approaches, pFISTA has the advantages of faster convergence speed, insensitivity to step size, and only needs to set one parameter.

Recently, convolutional neural network (CNN) has shown great potential in many fields [21-25]. However, the lack of understandable architecture hinders its medical application [26, 27]. Unrolling optimization methods into deep neural networks is a novel way to utilize the information of numerous databases and enhance network interpretability [12-14].

pFISTA-Net is an unrolled deep neural network for parallel MRI reconstruction [20]. In pFISTA-Net, manual-craft sparsity transform is replaced by learnable convolutional filters, making it more robust for reconstructing images in different scenarios. The residual structure is employed to improve the learning capabilities of the network. Experimental results show that pFISTA-Net has lower reconstruction error than state-of-the-art approaches, and is robust to different sampling patterns.

As the outstanding approaches for MRI reconstruction, pFISTA and pFISTA-Net are cited widely and compared by

This work was supported in part by National Natural Science Foundation of China (61971361, 61871341, and 61811530021), National Key R&D Program of China (2017YFC0108703), Science and Technology Program of Xiamen (3502Z20183053), Xiamen University Nanqiang Outstanding Talents Program. (\*Corresponding author: Xiaobo Qu with e-mail: quxiaobo@xmu.edu.cn).

Yirong Zhou, Chen Qian, Jian Wang, Zi Wang and Xiaobo Qu are with Biomedical Intelligent Cloud R&D Center, Department of Electronic Science, National Institute for Data Science in Health and Medicine, Xiamen University, Xiamen 361005, China.

Biao Qu is with Department of Instrumental and Electrical Engineering, Xiamen University, Xiamen 361005, China

Yi Guo and Di Guo are with School of Computer and Information Engineering, Xiamen University of Technology, Xiamen 361024, China.

Yongfu You is with China Mobile Group, Xiamen 361005, China; Biomedical Intelligent Cloud R&D Center, School of Electronic Science and Engineering, Xiamen University, Xiamen 361005, China many followed works [10, 14, 28-30]. As shown in Figure 1, XCloud-pFISTA that integrates above-mentioned two approaches will make them easier for using by MRI researchers, and perform further medical image analysis.

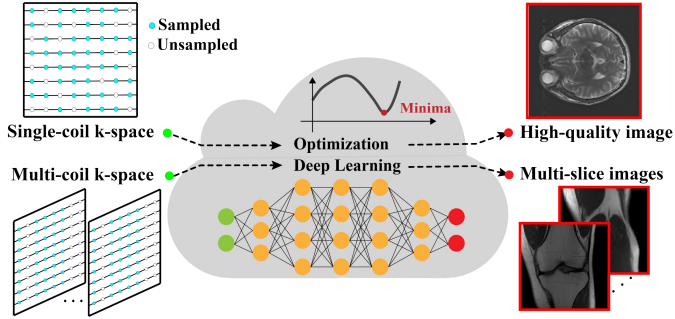

Figure 1. The schematic of XCloud-pFISTA.

## B. Cloud Computing Platform

Cloud computing is a kind of distributed computing, which is accessible, high-integrated, safe, and has virtualized hardware[31, 32]:

1. Cloud computing platform is accessible to users at all levels. Users can employ resource on the cloud through a user-friendly interface without coding.

2. Many MRI reconstruction approaches are developed in different software environment. Cloud computing platform provides a unified virtual environment for them.

3. As to the platform's security, we are developing a set of security mechanisms, including data encryption with Secure Sockets Layer (SSL) certificates and strict access policies. The platform is easy to be legally visited and could prevent data leakage effectively.

4. Thanks to hardware virtualization, the cloud platform is not limited by the number of computer kernels and data storage memory, making it possible to deploy many neural network models and big databases.

## C. System Architecture

Figure 2 shows the entire system architecture of XCloud-pFISTA. It adopts the browser/service (B/S) architecture, mainly consists of three parts: browser, service, and data access layer (DAL).

**Browser**: Users visit the website of XCloud-pFISTA through a browser without downloading or installing any software. The website has a user-friendly interface with some functional buttons on it. Each button is the visualization of a certain application programming interface (API), i.e., the call interface of service in the next part. Users can activate a certain API by pushing the corresponding button rather than coding, thus sending a service request to the next part.

**Service**: Requests will be transmitted by Nginx gateway following a complex and effective strategy. Here, pFISTA and pFISTA-Net are deployed on different web servers as core services. Web servers employ Google remote procedure call (gRPC) to communicate and network file system (NFS) to share data. These technologies ensure the high-effective and stability of the service part.

**DAL**: All data are stored on the DAL by an effective data storage strategy. MySQL is employed to store structural data, such as the user's name and password. And Redis is used to store frequently used data. Demo data and reconstruction data are stored in MongoDB, which is suitable for storing non-relational and large-size data.

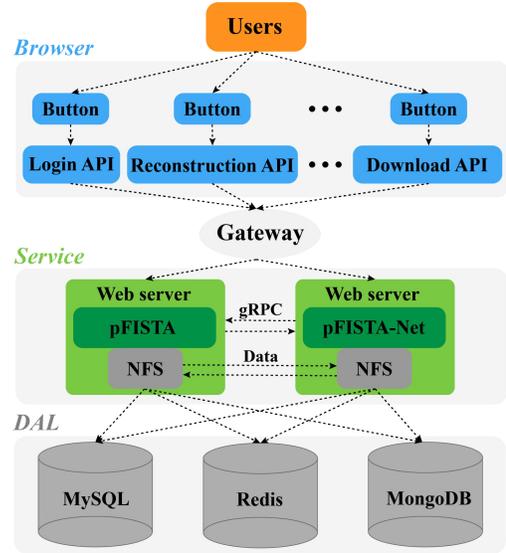

Figure 2. The system architecture of XCloud-pFISTA.

## III. RESULTS

### A. XCloud-pFISTA Workflow

Up to now, XCloud-pFISTA supports single-coil and multi-coil MRI reconstruction, and is open access at http://36.134.50.123:2345 (Test account: EMBC, Password: EMBC2021). Notably, we decide not to open up the registration in the peer review. The user manual and demo data are provided on the cloud for the quick try. Besides, you can also upload your own data and process them easily and quickly. Figure 3 shows the whole workflow of the usage of XCloud-pFISTA. It mainly includes five steps:

1. Pre-process k-space data into a required format before uploading them to the cloud platform.

2. Login XCloud-pFISTA through the test account and password.

3. Select pFISTA or pFISTA-Net for image reconstruction, according to the undersampled data (single-coil or multi-coil data).

4. After uploading all required files, click the submit button. Then users can download the reconstruction results with a short wait.

5. (Optional) Delete uploaded data and reconstruction results permanently. XCloud-pFISTA never save any data without the consent of the users.

Figure 4 shows the user interface of XCloud-pFISTA. The menu is very clear, including homepage, pFISTA, pFISTA-Net, and demo data (Figure 4(a)). The crucial interfaces of pFISTA and pFISTA-Net are shown in Figure 4(b) and Figure 4(c), respectively.

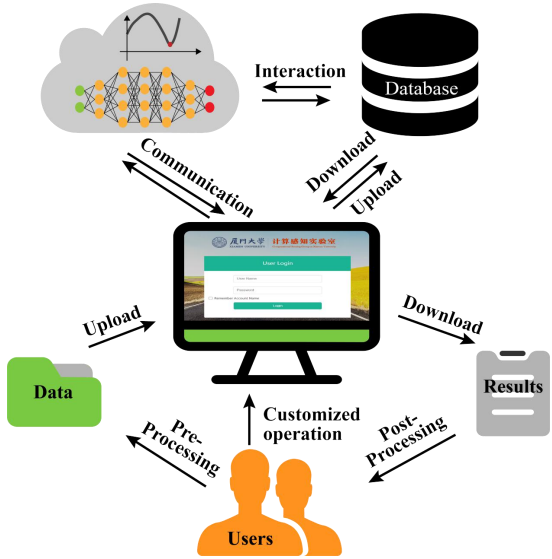

Figure 3. The workflow of XCloud-pFISTA for MRI reconstruction.

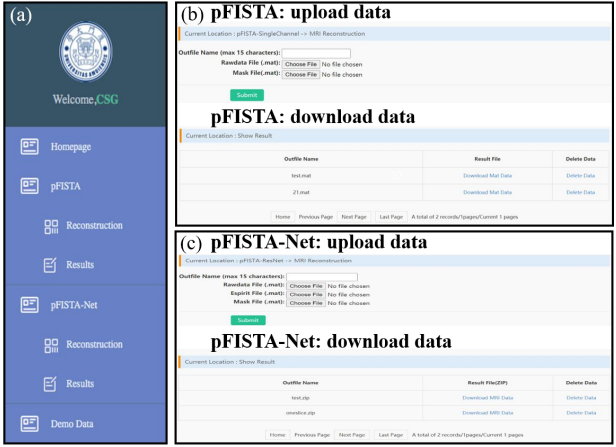

Figure 4. The user interface of XCloud-pFISTA. (a) Account information and menu. (b) pFISTA for single-coil data reconstruction. (c) pFISTA-Net for multi-coil data reconstruction.

### B. Reconstruction Results and Comparison

In order to verify the consistency of the reconstruction results and the performance of the cloud platform, we compare the cloud and local results. Both of pFISTA and pFISTA-Net are written in Python. The parameters used in pFISTA and pFISTA-Net are all consistent with original papers[19, 20]. Here, Local reconstruction is conducted on a Windows PC with 16 cores CPU, 128 GB RAM, and one NVIDIA GTX 850M. For reconstruction on cloud, we use the China Mobile e-cloud host with 16 cores CPU, 128 GB RAM, and two NVIDIA T4 GPUs. For quantitative comparison, we adopt the relative $l_2$ norm error (RLNE) defined as

$$\text{RLNE} := \| \mathbf{x} - \hat{\mathbf{x}} \|_2 / \| \mathbf{x} \|_2 \tag{1}$$

where $\mathbf{x}$ is the ground truth and $\hat{\mathbf{x}}$ is the reconstructed image.

For single-coil data, pFISTA is used to reconstruct the T2-weighted brain dataset, which is acquired from a healthy volunteer at a 3T Siemens Trio Tim MRI scanner with 32-coils using a T2-weighted turbo spin echo sequence (TR/TE = 6100/99 ms, FOV = 220×220 mm$^2$, slice thickness = 3mm). Results of pFISTA on local and cloud are shown in Figure 5.

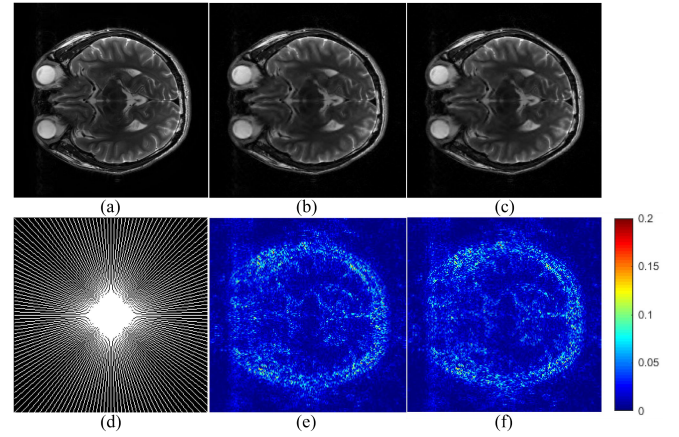

Figure 5. Reconstruction results of pFISTA on local and cloud. (a) is the ground truth. (b-c) are the reconstructed images of pFISTA on local and cloud, respectively. (d) is a 2D pseudo-radial mask (30%). (e-f) are error distributions corresponding to the above methods. The RLNEs of (b-c) are 0.102 and 0.102, respectively. The reconstruction time of (b-c) are 67.08 s and 31.99 s, respectively. Note: No parallel computing is used in reconstruction.

For multi-coil data, pFISTA-Net is used to reconstruct the knee dataset from a public knee dataset [33]: coronal proton-density (15 knee coils, TR/TE = 2750/27 ms, FOV = 320×320 mm$^2$, In-plane resolution = 0.49×0.44 mm$^2$). Results of pFISTA-Net on local and cloud are shown in Figure 6.

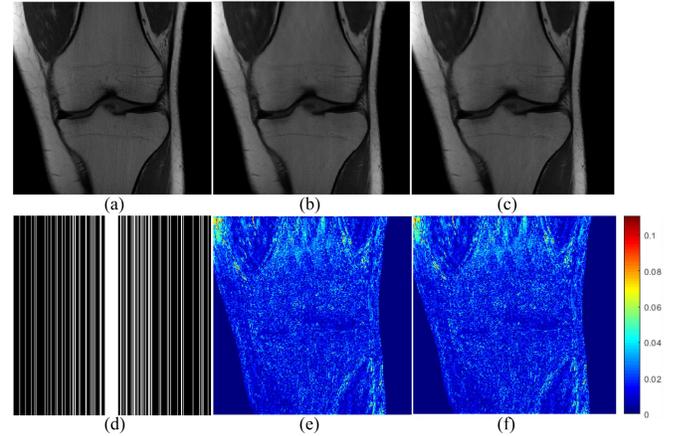

Figure 6. Reconstruction results of pFISTA-Net on local and cloud. (a) is the ground truth. (b-c) are the reconstructed images of pFISTA-Net on local and cloud, respectively. (d) is a Cartesian mask (25%). (e-f) are error distributions corresponding to the above methods. The RLNEs of (b-c) are 0.167 and 0.167, respectively. The reconstruction time of (b-c) are 10.33s and 2.59 s, respectively.

Figure 5 and Figure 6 show that XCloud-pFISTA obtains the same high-quality results as on local, while the reconstruction time is reduced remarkably. Thus, the results suggest that, XCloud-pFISTA is a reliable and high-performance cloud computing platform for accelerated MRI.

## IV. Conclusion and Outlook

In summary, we developed XCloud-pFISTA, a reliable, high-performance, and open access medical intelligence cloud for accelerated MRI. Two state-of-the-art approaches, pFISTA and pFISTA-Net, have been implemented on cloud for single-coil and multi-coil image reconstruction. We are constantly improving the XCloud-pFISTA and hope to provide an advanced cloud computing platform, which integrates numerous state-of-the-art approaches for MRI processing and analysis. Furthermore, we also plan to integrate artificial intelligence diagnostics on XCloud-pFISTA in the future.

## V. Acknowledgement

The authors thank China Telecom for providing cloud computing service support at initial. The authors also thank the research [33] for sharing a public knee dataset at: https://github.com/VLOGroup/mri-variationalnetwork. The authors appreciate the help of Jianzhong Lin and Xinlin Zhang for valuable suggestions.